\documentstyle[12pt]{article}
\def\T{T}
\def\P{P}
\def\Q{Q}
\def\half{{\textstyle\frac{1}{2}}}

\def\im{{\rm Im}}
\def\Res{\mathop{\rm Res}}
\newcommand{\ve}[1]{\hbox{\boldmath{$#1$}}}
\newcommand{\op}[1]{\hat{#1}}

\newcommand{\Ord}[1]{{\cal O}\left(#1\right)}
\newcommand{\Bra}[1]{\langle #1|}
\newcommand{\Ket}[1]{|#1\rangle}
\newcommand{\Braket}[2]{\langle #1|#2\rangle}
\newcommand{\bra}[1]{\{ #1|}
\newcommand{\ket}[1]{|#1\} }
\newcommand{\braket}[2]{\{ #1|#2\} }

\newtheorem{definition}{Definition}
\newtheorem{theorem}{Theorem}
\newtheorem{lemma}{Lemma}
\begin{document}
\vspace{1.0in}
\begin{center}
\Large
{\bf Quantum Surface of Section Method:\\
Decomposition of the Resolvent $(E-\op{H})^{-1}$\\}
\vspace{0.5in}
\large
Toma\v z Prosen \\
\normalsize
\vspace{0.3in}
Center for Applied Mathematics and Theoretical Physics,\\
University of Maribor, Krekova 2, SLO-62000 Maribor, Slovenia\\
\end{center}
\vspace{0.8in}

\noindent{\bf Abstract}
The paper presents exact surface of section reduction of quantum mechanics.
The main theoretical result is a decomposition of the energy-dependent
propagator $\op{G}(E) = (E - \op{H})^{-1}$ in terms of the propagators
which (also or exclusively) act in Hilbert space of complex-valued functions
over the configurational surface of section, which has one dimension less
than the original configuration
space. These energy-dependent quantum propagators from and/or onto the
configurational surface of section can be explicitly constructed as the
solutions of the first order nonlinear Riccati-like initial value problems.


\bigskip
\noindent Submitted to {\bf Zeitschrift f\" ur Physik B}

\newpage

\section*{I. INTRODUCTION}

In classical dynamics the concept of surface of section (SOS) was introduced
by Poincar\' e and it has been proved to be very useful ever since
\cite{LL83}. Dynamics of the Hamiltonian system with $f$ freedoms which is
a smooth diffeomorphism in $2f$-dimensional phase space can be shown to be
equivalent to discrete volume preserving mapping in $(2f-2)$-dimensional
(sub-)phase space. Every trajectory (which is confined to $(2f-1)$-dim energy
surface) typically crosses $(2f-2)$-dim intersection of $(2f-1)$-dim energy
surface and $(2f-1)$-dim SOS infinitely many times (if the system is bound
and the SOS is suitably chosen). The so called Poincar\' e map maps
one such crossing to the next one and it is area (volume) preserving.
Thus, for example, Hamiltonian systems of two freedoms can be reduced to area
preserving maps of a 2-dim plane.

There were few attempts to extend a concept of SOS reduction to quantum
mechanics but none have been completely successful yet. The quantum Poincar\' e
map would be an energy-dependent propagator which would map (not yet
firmly defined) amplitude distribution function over the $(f-1)$-dim projection
of $SOS$ to the configuration space to another. In bound systems, where no
trajectory can escape, the quantum Poincar\' e map should become unitary
in the semiclassical limit (It need not be unitary in nonsemiclassical regime).
Bogomolny \cite{B92} proposes semiclassical propagator for the SOS map which is
symmetric and approximately unitary (as $\hbar\rightarrow 0)$.
Bogomolny's map has a clear
classical interpretation in terms of the classical trajectories,
actions along them and corresponding monodromy matrices, but its
greatest deficiency is that it is essentially semiclassical, so it cannot
be viewed directly as an approximation of some exact quantum propagator.
Smilansky, Doron and Schanz \cite{DS92,SS93} propose scattering approach for
diagonalization of bound systems. Their method is almost exact (the error is of
order $\Ord{\exp(-{\rm const}/\hbar)}$ because they throw away the most of
atenuating (evanescent) modes with imaginary wave number in order to make the
basis for the $S$-matrix finite) but it is specialized for quantum billiards
(Laplacians with Dirichlet boundary conditions).
In fact for the billiards, Bogomolny`s method is a limiting case
(as $\hbar\rightarrow 0$) of the method of Smilansky and coworkers.
Recently, Gutzwiller \cite{G93} published the exact quantum surface of section
map for one system, namely, the free particle on a ``trombone'' attached to a
parallelogram. My approach can be
viewed as a generalization of the scattering approach to arbitrary
quantum Hamiltonian system whose kinetic energy perpendicular to SOS
is quadratic. Here I present a {\em local} version of the quantum surface of
section method and prove its consistency in fairly rigorous manner.
The central result is the proof of the decomposition formula
for the energy-dependent Green's function in terms of energy-dependent
propagators which act also (or exclusively, like my quantum Poincar\' e map)
in the Hilbert space of functions over the configurational surface of section.
Shortly after the present work had been completed I managed to develop a more
straightforward {\em global} approach \cite{P94a,P94b} in terms of quantum
mechanical scattering theory, although the present local theory gives some
interesting results which were not obtained in \cite{P94a,P94b}.
I have also applied the present method to calculate the eigenenergies
with the sequential number around 20 million in a nonintegrable system,
namely the so called {\em 2-dim semi-separable oscillator} \cite{P94c}
which dynamically has all the generic features even though it is
geometrically somewhat special. To the best of my knowledge this is the
only available method enabling to calculate such high-lying eigenstates,
which demonstrates not only the conceptual importance but also the
practical power of the present method.
After this work had been completed I have been informed about a related work
on the exact quantum SOS method of
$2-$dim Hamiltonians of the type $\ve{p}^2/2m + V(\ve{q})$
by Rouvinez and Smilansky \cite{RS94}.
\\\\
In the Section II we will introduce convenient notation
together with all the necessary definitions. Then the decomposition formula
for the resolvent of the energy-dependent Green's function in terms of
propagators which propagate from and/or onto the SOS will be presented and a
comment on its interpretation will be given.
In Section III the proof of the decomposition formula will be illustrated for
the systems with single degree of freedom (1-dim configuration space).
The proof of the general decomposition formula will be given in
Section IV. Additional discussion on the interpretation of the quantum surface
of section method and conclusions are found in section V.

\section*{II. DECOMPOSITION OF THE RESOLVENT}

First I will introduce convenient notation for posing the problem.
Vectors in $f$-dim configuration space manifold ${\cal C}$ (usually
${\cal C} = \Re^f$) will be denoted by $\ve{q}$ and vectors from the
corresponding conjugate momentum space will be written as $\ve{p}$.
The discussion will be limited only to the case where the
SOS is perpendicular to the configuration space (CS),
${\rm SOS} = \{(\ve{q},\ve{p}); s(\ve{q}) = 0\}$. More general
cases can generally be transformed to the upper one by the use of the canonical
phase space transformations. Thus, we are considering only configurational
surface of section ${\cal S}_0 = \{\ve{q}\in {\cal C}; s(\ve{q}) = 0\} \subset
{\cal C}$. Assume that the topology of CS is simply connected
and that function $s$ can be globalized in a way that all members of the
{\em family of surfaces of section} ${\cal S}_y = \{\ve{q}\in {\cal C};
s(\ve{q}) = y\}$ are topologically equivalent. Then the CS
can be written as a Cartesian product ${\cal C} = {\cal S}\times
[y_\downarrow,y_\uparrow]$, inducing the separation of coordinates
$\ve{q} = (\ve{x},y),\;\ve{p} = (\ve{p}_x,p_y)$. Therefore one may write
${\cal S}_y = ({\cal S},y)$. For the most useful
example of Euclidean configuration space $\subseteq \Re^f$ with flat, mutually
parallel surfaces of section this is already done by considering
$(\ve{x},y)$ as coordinates in Cartesian coordinate system where $y$-axis is
oriented perpendicularly to SOS. The boundary surfaces
$(\ve{x},y_{\uparrow\downarrow})$ can be either at finite or infinite distance
$y_{\uparrow\downarrow}\in \Re\cup\{-\infty,+\infty\}$.
Every surface of section cuts the CS in two pieces which will be referred to as
upper and lower and denoted by the value of the index $\sigma=\uparrow,
\downarrow$. In arithmetic expressions, the arrows will have the following
values, $\uparrow = +1,\downarrow = -1.$
So, we will be
considering not just one, but a whole family of surfaces of section at a time,
where subvector $\ve{x}$ will denote the position inside SOS (parallel
coordinates) and $y$ will label the given SOS
(perpendicular coordinate). This will be
useful since it will turn out that the quantum SOS-SOS propagator (as will be
defined below) for ${\cal S}_y$ uniquely determine such propagator
for ${\cal S}_{y+dy}$ in nonlinear first order (Riccati-like) differential
equation setting.

My theory presented in this paper will apply to quite general class
of bound Hamiltonians (with possible generalizations to non-bound scattering
problems) whose kinetic energy is quadratic at least perpendicularly to
SOS
\begin{equation}
H = \frac{1}{2m} p_y^2 + H^\prime (\ve{p}_x,\ve{x},y).
\label{eq:clH}
\end{equation}
The first term characterizes the perpendicular motion among surfaces of
section and the second term captures the dynamics inside SOS.
For some Hamiltonians which are not originally of this form this can be
achieved by a proper choice of the SOS, i.e. the {\em coordinate system}
$(\ve{x},y)$. The parallel phase space coordinates $(\ve{x},
\ve{p}_{\ve{x}})$ may not be necessarily of the Weyl-type but they may
belong to arbitrary Lie group manifold as will become more clear
when the theory will be developed (see discussion in section V).
\\\\
In quantum mechanics, the observables are represented by self-adjoint
operators in a Hilbert space ${\cal H}$ of complex-valued functions
$\Psi(\ve{q})$ over the CS ${\cal C}$ which obey
boundary conditions $\Psi(\ve{x},y_{\uparrow\downarrow})=0$ and
$\Psi(\ve{x}\in\partial{\cal S}\;{\rm or}\;|\ve{x}|\rightarrow\infty,y)=0$
and have finite norm $\int_{\cal C} d\ve{q}|\Psi(\ve{q})|^2 < \infty$. We will
use Dirac's notation. Pure state of a
physical system can be represented by a vector --- {\em ket} $\Ket{\Psi}$
which can be expanded in a convenient complete set of basis vectors, e.g.
position eigenvectors $\Ket{\ve{q}}=\Ket{\ve{x},y},\;
\Ket{\Psi} = \int_{\cal C}d\ve{q}\Ket{\ve{q}}\Braket{\ve{q}}{\Psi}=
\int_{\cal C}d\ve{q}\Psi(\ve{q})\Ket{\ve{q}}$
(in a symbolic sense, since $\Ket{\ve{q}}$ are not proper vectors,
but such expansions are still meaningful iff $\Psi(\ve{q})$ is
square integrable i.e. $L^2$-function).
Every ket $\Ket{\Psi}\in{\cal H}$ has a corresponding vector from the
dual Hilbert space ${\cal H}^\prime$, that is {\em bra}
$\Bra{\Psi}\in{\cal H}^\prime,\; \Braket{\Psi}{\ve{q}} =
\Braket{\ve{q}}{\Psi}^*$. Now, fix a given SOS, $y = {\rm const}$. Operators
$\op{\ve{x}}$ and $\op{\ve{p}}_{\ve{x}}$ can be viewed as acting on functions
$\psi(\ve{x})$ of $\ve{x}$ only and therefore operating in some other, much
smaller Hilbert space of square-integrable complex-valued functions over a
given SOS ${\cal S}_y$. A family of such SOS-Hilbert spaces parametrized by
$y$ will be denoted by ${\cal L}_y$. Vectors in a SOS-Hilbert space ${\cal
L}_y$
will be denoted by $\ket{\psi}_y$ (where subscript $y$ will be
omitted if possible). Eigenvectors of SOS position operators $\op{\ve{x}},\;
\op{\ve{x}}\ket{\ve{x}^\prime} = \ve{x}^\prime\ket{\ve{x}^\prime}$
can provide a useful choice for the complete set of basis vectors of
${\cal L}_y$. $\braket{\ve{x}}{\psi} = \braket{\psi}{\ve{x}}^*$ can be
interpreted as a quantum mechanical amplitude for a trajectory of a system
being
in quantum SOS-state $\ket{\psi}$ to cross SOS ${\cal S}_y$ at the point
$\ve{x}$. Since surfaces ${\cal S}_y = ({\cal S},y)$ are all topologically
equivalent, the SOS-Hilbert spaces ${\cal L}_y$ are also all isomorphic, i.e.
equivalent to some abstract SOS-Hilbert space ${\cal L}$. Isomorphism
$\op{I}_{y^\prime y}\in Lin({\cal L}_y,{\cal L}_{y^\prime})$ (where
$Lin({\cal A},{\cal B})$ denotes the set of all linear mappings from
Hilbert space ${\cal A}$ to Hilbert space ${\cal B}$) is being
provided by universal definition of the (symbolic) SOS-position eigenstates
$\ket{\ve{x}}_y \equiv \ket{\ve{x}}$
\begin{equation}
\op{I}_{y^\prime y} = \int\limits_{\cal S}d\ve{x}\,\ket{\ve{x}}_{y^\prime}\,
{_y\bra{\ve{x}}},\label{eq:iso}
\end{equation}
which is the well known decomposition of identity of ${\cal L}_y$ for
$y=y^\prime$.
The quantum Hamiltonian (like the classical one (\ref{eq:clH})) can be
written as a sum of two terms
\begin{equation}
\op{H} = -\frac{\hbar^2}{2m}\partial_y^2 + \op{H}^\prime,\quad
\op{H}^\prime = H^\prime(\op{\ve{p}}_x,\op{\ve{x}},\op{y}) =
H^\prime(-i\hbar\partial_{\ve{x}},\ve{x},y),
\label{eq:qH}
\end{equation}
where the eigenstates of the {\em inside Hamiltonian} $\op{H}^\prime$
restricted to the SOS-Hilbert space ${\cal L}_y$ (treating $y$ as a
parameter and not an operator), $\ket{n}_y\in{\cal L}_y$ called
SOS-eigenmodes
\begin{equation}
\op{H}^\prime\vert_{{\cal L}_y}\ket{n}_y = E^\prime_{n}(y)\ket{n}_y
\label{eq:eigenmodes}
\end{equation}
provide very useful (countable $n=1,2,\ldots$) complete basis of
${\cal L}_y$ as will soon become clear.

\begin{definition}
\label{def:H}
Let us define two families of Hilbert subspaces ${\cal H}^\sigma_{E y_0}
\subset {\cal H}, \sigma =\uparrow,\downarrow$ of $L^2$-functions which
have definite energy, i.e. they satisfy Schr\" odinger equation,
above/below a given SOS ${\cal S}_{y_0}$
and are zero below/above ${\cal S}_{y_0}$
\begin{equation}
\begin{array}{lll}
{\cal H}^\sigma_{E y_0} = \{ \Ket{\Psi}\in{\cal H};\;
&\op{H}\Psi(\ve{x},y) = E\Psi(\ve{x},y) &{\rm if}\;\;\sigma y \ge \sigma y_0,\\
&\Psi(\ve{x},y) = 0,& {\rm if}\;\; \sigma y < \sigma y_0 \}.
\end{array}
\label{eq:Hilb}
\end{equation}
\end{definition}
The values of $\Ket{\Psi}\in {\cal H}^\sigma_{E y_0}$ on the SOS
${\cal S}_{y_0}$, $\Psi(\ve{x},y_0)$
are quite arbitrary \footnote{
Of course, they should decay fast enough when $|\ve{x}| \rightarrow \infty$ in
case when SOS is infinite.} and represent a boundary condition
for the time-independent Schr\" odinger equation which together with
the remaining boundary condition $\Psi(\ve{x},y_\sigma) = 0$ determines
$\Ket{\Psi}$ uniquely. This suggests that the spaces ${\cal L}_{y_0}$ and
${\cal H}^\sigma_{E y_0}$ might be of the same size (elements of both
are uniquely determined by the values of certain function on the $(f-1)$-dim
SOS) and motivates the search for the connection between them.
In order to achieve this we will use SOS-eigenmode (\ref{eq:eigenmodes})
expansion of a state $\Ket{\Psi}\in{\cal H}^\sigma_{E y_0}$ near the surface
${\cal S}_{y_0}, y = y_0 + \sigma\epsilon, \epsilon\ge 0$,
\begin{equation}
\Psi(\ve{x},y) = \sum\limits_n \braket{\ve{x}}{n}_{y_0}
\sqrt{\frac{-im}{\hbar^2 k_n(E,y_0)}}
\left[ c^{\rm out}_n e^{i\sigma k_n(E,y_0)(y-y_0)} +
       c^{\rm onto}_n e^{-i\sigma k_n(E,y_0)(y-y_0)}
\right] + \Ord{\epsilon^2},
\label{eq:exp1}
\end{equation}
which is exact iff $\partial_y k_n(E,y) \equiv 0$, i.e. inside hamiltonian
$\op{H}^\prime$ does not depend on the ``parameter'' $y$.
Throughout this paper we will use the following definition of the complex
square root $\sqrt{z} = \sqrt{\half(|z| + {\rm Re}\,z)} +
i{\rm sgn}\,{\rm Im}\,z \sqrt{\half(|z| - {\rm Re}\,z)}$
which is a complex analytic function in upper half-plane $y\ge 0$.
The accuracy of the ansatz (\ref{eq:exp1}) cannot be extended beyond the
second order without extending the set of coefficients
and differentiating quantities which depend on $y_0$,
but this is quite enough, since the present form (\ref{eq:exp1}) correctly
reproduces normal derivative $\partial_y \Psi(\ve{x},y)$ and therefore
also the probability currents through the SOS. The wavenumbers
\begin{equation}
k_n(E,y_0) = \sqrt{\frac{2m}{\hbar^2}(E - E_n(y_0))},
\label{eq:wavenumb}
\end{equation}
which correspond to SOS-eigenmodes $\ket{n}_{y_0}$, assign the rest of the
energy $E-E_n(y_0)$ to the perpendicular motion providing all terms in
expansion (\ref{eq:exp1}) with the constant and correct value of the
{\em effective} energy. The evanescent modes, for which $E_n(y_0) > E$, have
imaginary wavenumbers.
The coefficients $c^{\rm out}_n,c^{\rm onto}_n$ can be interpreted as the
probability currents for SOS-eigenmodes to propagate out from the SOS and back
onto the SOS, respectively. But only half of them are independent since the
half of their nondegenerate linear combinations are arbitrary
\begin{equation}
c^{\rm out}_n + c^{\rm onto}_n = \sqrt{\frac{\hbar^2 k_n(E,y_0)}{-im}}
\int\limits_{\cal S}d\ve{x}\Psi(\ve{x},y_0)\braket{n}{\ve{x}}
\label{eq:coef1}
\end{equation}
since boundary condition $\Psi(\ve{x},y_0)$ is arbitrary and
the other half of their linear combinations are fixed
\begin{equation}
c^{\rm out}_n - c^{\rm onto}_n = \sigma\sqrt{\frac{-i\hbar^2}{m k_n(E,y_0)}}
\int\limits_{\cal S}d\ve{x}\partial_y \Psi(\ve{x},y_0)\braket{n}{\ve{x}}
\label{eq:coef2}
\end{equation}
since the normal derivative $\partial_y \Psi(\ve{x},y_0)$ is determined
by the solution of the Schr\" odinger equation $\op{H}\Psi = E\Psi$ with
boundary conditions $\Psi(\ve{x},y_0)$ and $\Psi(\ve{x},y_\sigma) = 0$.
We choose the coefficients $c^{\rm out}_n$ to be arbitrary so the remaining
coefficients $c^{\rm onto}_n$ are uniquely determined by $c^{\rm out}_n$,
and make the following definition.

\begin{definition}
\label{def:T}
Take an arbitrary out-going quantum SOS-state $\ket{\psi^{\rm out}}$ with the
eigenmode expansion $\ket{\psi^{\rm out}} = \sum\limits_n c^{\rm out}_n \ket{n}
$. Then the propagator from SOS back to SOS (above/below
SOS if $\sigma=\uparrow\downarrow$, respectively) can be defined
\begin{equation}
\op{\T}_\sigma (E,y_0) \in Lin({\cal L}_{y_0},{\cal L}_{y_0})
\end{equation}
and the on-going quantum SOS-state $\ket{\psi^{\rm onto}} = \sum\limits_n
c^{\rm onto}_n \ket{n}$ can be written as
\begin{equation}
\ket{\psi^{\rm onto}} = \op{\T}_\sigma (E,y_0) \ket{\psi^{\rm out}}.
\end{equation}
Generally,
\begin{equation}
\op{\T}_\sigma(E,y_0) = \sum\limits_{n,n^\prime}
c^{\rm onto}_n\vert_{(c^{\rm out}_l = \delta_{l n^\prime})}
\ket{n}\bra{n^\prime}.
\end{equation}
\end{definition}
Let us join all the information about the SOS-eigenmodes (\ref{eq:eigenmodes})
and their wavenumbers in the definition of the waveoperator $\op{K}(E,y)$.
\begin{definition}
\label{def:K}
Linear operator $\op{K}(E,y)\in Lin({\cal L}_y)$ is by definition an operator
whose spectrum is just the set of wavenumbers $\{k_n(E,y)\}$ with
SOS-eigenvectors $\ket{n}_y$
\begin{equation}
\op{K}(E,y) = \sum\limits_n k_n(E,y) \ket{n}_y\,{_y\bra{n}} =
\sqrt{\frac{2m}{\hbar^2}\left(E - \op{H}^\prime\vert_{{\cal L}_y}\right)}.
\end{equation}
\end{definition}
With aid of this new definition and the notation of Definition \ref{def:T}
one can write SOS-eigenmode expansion (\ref{eq:exp1}) more compactly
\footnote{
The SOS-SOS propagator $\op{T}_\sigma(E,y_0)$ can be viewed also as a
scattering operator of the following scattering problem which is obtained
from the orginal bound Hamiltonian
$\op{p}_y^2/2m + H^\prime(\op{p}_{\ve{x}},\ve{x},y)$
by substituting it in one part of CS,
$\sigma y \le \sigma y_0$, by the $y-$flat waveguide with the Hamiltonian
$\op{p}_y^2/2m + H^\prime(\op{p}_{\ve{x}},\ve{x},y_0)$, for $\sigma y\le
\sigma y_0$.}
\begin{eqnarray}
\Psi(\ve{x},y) &=& \frac{\sqrt{-im}}{\hbar}\bra{\ve{x}}\op{K}^{-1/2}(E,y_0)
\left[e^{i\sigma\op{K}(E,y_0)(y-y_0)} +
e^{-i\sigma\op{K}(E,y_0)(y-y_0)}\op{T}_\sigma(E,y_0)\right]
\ket{\psi^{\rm out}} \nonumber \\
&+& \Ord{|y-y_0|^2}
\label{eq:compexp}
\end{eqnarray}
\begin{lemma}
\label{lem:L1}
Expansion (\ref{eq:compexp}) holds for all states $\Ket{\Psi}$
from all spaces ${\cal H}^\sigma_{E y_1}$, such that
$\sigma y_1 \le \sigma y_0$.
\end{lemma}
The proof follows solely from the definition \ref{def:H} of the
spaces ${\cal H}^\sigma_{E y}$, since all ${\cal H}^\sigma_{E y_1}$ are
equivalent on the domain $\{(\ve{x},y);\ve{x}\in{\cal S},\sigma y_0
\le \sigma y \le \sigma y_\sigma\}$, for $\sigma y_1 \le \sigma y_0$.
\begin{definition}
\label{def:PQ}
Let $\op{\P}_\sigma(E,y_0)$ and $\op{Q}_\sigma(E,y_0)$ be
the quantum propagators from upper/lower ($\sigma=\uparrow\downarrow$) part
of CS ${\cal C}$ to SOS ${\cal S}_{y_0}$ and vice-versa
\begin{eqnarray}
\op{\Q}_\sigma(E,y_0) &\in& Lin({\cal L}_{y_0},{\cal H}) \\
\op{\P}_\sigma(E,y_0) &\in& Lin({\cal H},{\cal L}_{y_0})
\end{eqnarray}
such that for arbitrary states
$\ket{\psi^{\rm out}}\in {\cal L}_{y_0},
 \bra{\psi^{\rm onto}}\in {\cal L}^\prime_{y_0}$
\begin{eqnarray}
\op{Q}_\sigma(E,y_0)\ket{\psi^{\rm out}} &\in& {\cal H}^\sigma_{E y_0} \\
\bra{\psi^{\rm onto}}\op{P}_\sigma(E,y_0) &\in&
{\cal H}^{\sigma\prime}_{E^* y_0}
\label{eq:PQ1}
\end{eqnarray}
which are uniquely determined by the boundary condition on the SOS
${\cal S}_{y_0}$
\begin{eqnarray}
\Bra{\ve{x},y_0}\op{Q}_\sigma(E,y_0)\ket{\psi^{\rm out}} &=&
\frac{\sqrt{-im}}{\hbar}\bra{\ve{x}}\op{K}^{-1/2}(E,y_0)(1 +
\op{T}_\sigma(E,y_0))\ket{\psi^{\rm out}} \label{eq:valQ}\\
\bra{\psi^{\rm onto}}\op{P}_\sigma(E,y_0)\Ket{\ve{x},y_0} &=&
\frac{\sqrt{-im}}{\hbar}\bra{\psi^{\rm onto}}(1 + \op{T}_\sigma(E,y_0))
\op{K}^{-1/2}(E,y_0)\ket{\ve{x}} \label{eq:valP}
\end{eqnarray}
\end{definition}
Here, perhaps, a remark is in order:
Conjugated energy $E^*$ was used in $(\ref{eq:PQ1})$ in order to make
propagator $\op{P}_\sigma(E,y)$ complex analytic function of $E$ rather than
$E^*$.

Now all the necessary tools are prepared to state the main result of
this paper:
\begin{theorem}
\label{the:main}
The energy-dependent quantum propagator (i.e. the resolvent of the Hamiltonian)
$\op{G}(E) = (E - \op{H})^{-1}$ can be decomposed in terms of the
CS-CS propagator --- with {\em no intersection with the} SOS ${\cal S}_y$ ---
$\op{G}_0(E,y)$, CS-SOS propagator $\op{\P}_\sigma(E,y)$, SOS-CS
propagator $\op{\Q}_\sigma(E,y)$, and SOS-SOS propagator $\op{\T}_\sigma(E,y)$
\begin{eqnarray}
\op{G}(E) = \op{G}_0(E,y) &+& \sum\limits_\sigma \op{\Q}_\sigma(E,y)
(1-\op{\T}_{-\sigma}(E,y)\op{\T}_{\sigma}(E,y))^{-1} \op{\P}_{-\sigma}(E,y)
\nonumber \\
&+& \sum\limits_\sigma \op{\Q}_{\sigma}(E,y)
(1-\op{\T}_{-\sigma}(E,y)\op{\T}_{\sigma}(E,y))^{-1}
\op{T}_{-\sigma}(E,y)\op{\P}_{\sigma}(E,y),
\label{eq:decomp}
\end{eqnarray}
and all poles of the Green's function $\op{G}(E)$
are in one-to-one correspondence with the singularities of
$(1-\op{\T}_{\sigma}(E,y)\op{\T}_{-\sigma}(E,y))^{-1}$
where the four propagators $\op{\P}_\sigma(E,y),\op{\Q}_\sigma(E,y),
\op{\T}_\sigma(E,y),$ and $\op{G}_0(E,y)$ are analytic.
\footnote{The latter propagators are non-analytic e.g. at thresholds for
opening of new modes $E^\prime_n$ where $k_n(E^\prime_n,y) = 0$.}
\end{theorem}
The propagator $\op{G}_0(E,y)$ can be calculated from (\ref{eq:decomp}) if the
Green's function $\op{G}(E)$ is known explicitly, but it is less
important, since its contribution becomes negligible when the energy $E$
approaches an eigenenergy (pole). Note that $\Bra{\ve{q}^\prime}
\op{G}_0(E,y)\Ket{\ve{q}}$ is the probability amplitude to propagate from
point $\ve{q}$ to point $\ve{q}^\prime$ at energy $E$ and without hitting the
SOS ${\cal S}_y$ in between, so it is equal to zero if initial and final point,
$\ve{q}$ and $\ve{q}^\prime$, lie on different sides of the SOS ${\cal S}_y$.
The decomposition formula (\ref{eq:decomp}) has
a very strong physical interpretation. In order to see this most clearly we
omit the arguments $(E,y)$ and formally
expand $(1-\op{T}_\sigma\op{T}_{-\sigma})^{-1}$ in terms of the geometric
series
\begin{eqnarray}
\op{G} = \op{G}_0 &+& \op{\Q}_\downarrow\op{\P}_\uparrow +
\op{\Q}_\downarrow\op{T}_\uparrow\op{T}_\downarrow\op{\P}_\uparrow +
\op{\Q}_\downarrow\op{T}_\uparrow\op{T}_\downarrow
\op{T}_\uparrow\op{T}_\downarrow\op{\P}_\uparrow + \ldots
\nonumber \\
&+& \op{\Q}_\uparrow\op{\P}_\downarrow +
\op{\Q}_\uparrow\op{T}_\downarrow\op{T}_\uparrow\op{\P}_\downarrow +
\op{\Q}_\uparrow\op{T}_\downarrow\op{T}_\uparrow
\op{T}_\downarrow\op{T}_\uparrow\op{\P}_\downarrow + \ldots
\nonumber \\
&+& \op{\Q}_\uparrow\op{\T}_\downarrow\op{\P}_\uparrow +
\op{\Q}_\uparrow\op{\T}_\downarrow\op{T}_\uparrow\op{T}_\downarrow
\op{\P}_\uparrow +
\op{\Q}_\uparrow\op{\T}_\downarrow\op{T}_\uparrow\op{T}_\downarrow
\op{T}_\uparrow\op{T}_\downarrow\op{\P}_\uparrow + \ldots
\nonumber \\
&+& \op{\Q}_\downarrow\op{\T}_\uparrow\op{\P}_\downarrow +
\op{\Q}_\downarrow\op{\T}_\uparrow\op{T}_\downarrow\op{T}_\uparrow
\op{\P}_\downarrow +
\op{\Q}_\downarrow\op{\T}_\uparrow\op{T}_\downarrow\op{T}_\uparrow
\op{T}_\downarrow\op{T}_\uparrow\op{\P}_\downarrow + \ldots
\label{eq:decomp2}
\end{eqnarray}
The arguments $(E,y)$ will also be omitted to increase transparency
in the rest of this paper whenever this will cause absolutely no confusion.
The meaning of this expression (\ref{eq:decomp2}) when put in a sandwich
between $\Bra{\ve{q}^\prime}$ and $\Ket{\ve{q}}$ is that the {\em probability
amplitude to propagate from point $\ve{q}$ to $\ve{q}^\prime$ at energy $E$}
(PAPPE) is a sum of: (i) PAPPE without hitting the SOS ($\op{G}_0$ term) plus
(ii) PAPPE with single intersection with SOS ($\op{\Q}\op{\P}$ terms) plus
(iii) PAPPE with double intersection with SOS ($\op{\Q}\op{\T}\op{\P}$ terms)
plus $\ldots$ The alternating symbols $\uparrow\downarrow$ in a compositum like
$\op{\Q}_\downarrow\op{\T}_\uparrow\op{\T}_\downarrow\op{\P}_\uparrow$
(see Figure 1) mean that
the system should be on different sides of the SOS before and after it crosses
the SOS, since typical quantum trajectory (in the sense of summing over paths)
is continuous. The proof of the statements made in this section will be given
in
Section IV. A more lengthy discussion on the interpretation can be found
in section V.

\section*{III. ONE-DIMENSIONAL CASE}

First we shall prove the Theorem \ref{the:main} for the special case of
systems with a one degree of freedom $f=1$. Thus the main idea of the
proof will be more transparent and the reader will follow the next
section, which is a natural continuation of the previous one, more easily.
Ozorio de Almeida \cite{A94} published a similar study of one-dimensional
systems in the context of exact quantum SOS method for {\em separable systems}.

Now the configuration space is just the interval
(or the whole real axis) ${\cal C} = [y_\downarrow,y_\uparrow]$ and the
surfaces
of section ${\cal S}_{y_0}$ are single points $y = y_0$. Thus the
SOS-Hilbert spaces ${\cal L}_{y_0}$ are one-dimensional with one
normalized basis vector $\ket{1}$ and the
SOS-SOS propagator is just a simple complex-valued
function $T_\sigma(E,y_0) = \bra{1}\op{T}_\sigma(E,y_0)\ket{1}$
which can be also represented with the
phase shift $\delta_\sigma(E,y_0)$ between the outgoing and the incoming wave
in the local expansion (analogous to (\ref{eq:exp1},\ref{eq:compexp})) of the
(real) wavefunction $\Ket{\Psi^\sigma_{E y_0}}\in{\cal H}^\sigma_{E y_0}$
\begin{eqnarray}
\Psi^\sigma_{E y_0}(y) &=& \sqrt{\frac{4m}{\hbar^2 k(E,y_0)}}
\cos\left(k(E,y_0)(y-y_0) + \half\delta_\sigma(E,y_0)\right) +
\Ord{\epsilon^2}, \label{eq:schr} \\
T_\sigma(E,y_0) &=& e^{-i\sigma\delta_\sigma(E,y_0)},
\end{eqnarray}
where $y = y_0 + \sigma\epsilon,\,\epsilon\ge 0$.
The propagators onto/from the SOS $\P_\sigma(E,y_0,y) =
\bra{1}\op{P}_\sigma(E,y_0)\Ket{y}$ and $\Q_\sigma(E,y,y_0) =
\Bra{y}\op{Q}_\sigma(E,y_0)\ket{1}$
are now unique solutions of the Schr\" odinger equations
\begin{equation}
\begin{array}{l}
(\partial_y^2 + k^2(E,y)) \P_\sigma(E,y_0,y) = 0\\
(\partial_y^2 + k^2(E,y)) \Q_\sigma(E,y,y_0) = 0
\end{array}
\quad
k^2(E,y) = \frac{2m}{\hbar^2}(E - V(y))
\end{equation}
with boundary conditions
\begin{eqnarray}
\Q_\sigma(E,y_0,y_0) &=& \P_\sigma(E,y_0,y_0) =
\sqrt{\frac{-4im}{\hbar^2 k(E,y_0)}}\cos\left(\half\delta_\sigma(E,y_0)\right)
e^{-i\sigma\half\delta_\sigma(E,y_0)} \\
\Q_\sigma(E,y_\sigma,y_0) &=& \P_\sigma(E,y_0,y_\sigma) = 0.
\end{eqnarray}
It is convenient to express the SOS-CS-SOS propagators in terms of
the standardized (real) solution of the Schr\" odinger equation
\begin{equation}
\Q_\sigma(E,y_0,y) = \P_\sigma(E,y,y_0) = \Psi^\sigma_{E y_0}(y)
e^{-i\sigma\half\delta(E,y_0)-i{\textstyle\frac{\pi}{4}}}.
\end{equation}
Let the matrix elements of the Green's functions be denoted by
$G(E,y^\prime,y) = \Bra{y^\prime}\op{G}(E)\Ket{y}$,
$G_0(E,y_0,y^\prime,y) = \Bra{y^\prime}\op{G}_0(E,y_0)\Ket{y}$.
The decomposition formula in one dimension can now be written as
\begin{eqnarray}
G(E,y^\prime,y) &=& G_0(E,y_0,y^\prime,y) + \nonumber \\
&+& \Q_\downarrow (E,y^\prime,y_0)
(1 - \T_\uparrow\T_\downarrow)^{-1}\P_\uparrow(E,y_0,y) +
\nonumber \\
&+& \Q_\uparrow (E,y^\prime,y_0)
(1 - \T_\downarrow\T_\uparrow)^{-1}\P_\downarrow(E,y_0,y) +
\nonumber \\
&+& \Q_\uparrow (E,y^\prime,y_0)
(1 - \T_\downarrow\T_\uparrow)^{-1}\T_\downarrow\P_\uparrow(E,y_0,y) +
\nonumber \\
&+& \Q_\downarrow (E,y^\prime,y_0)
(1 - \T_\uparrow\T_\downarrow)^{-1}\T_\uparrow\P_\downarrow(E,y_0,y) =
\label{eq:decomp1d1} \\
&=& G_0(E,y^\prime,y) -
\frac{\sum\limits_\sigma\left(
\Psi^\sigma_{E y_0}(y^\prime)\Psi^{-\sigma}_{E y_0}(y) +
\Psi^\sigma_{E y_0}(y^\prime)\Psi^\sigma_{E y_0}(y)e^{-i\half\delta}
\right)}{2\sin\left(\half\delta\right)}
\label{eq:decomp1d2}
\end{eqnarray}
where $\delta = \delta_\uparrow(E,y_0) - \delta_\downarrow(E,y_0)$.
There are then two points to be proved:
\begin{itemize}
\item {\bf Point 1}
All the poles of the Green function $G(E,y^\prime,y)$ in complex energy $E$
plane should come from singularities of $(1-\T_\uparrow(E,y_0)
\T_\downarrow(E,y_0))^{-1}$ and vice versa.
\item {\bf Point 2} The residuum of the decomposed
Green function $G(E,y^\prime,y)$ at the pole --- eigenenergy $E_0$ should be
equal to $\Psi(y)\Psi(y^\prime)$ where $\Psi(y)$ is a {\em normalized} real
eigenfunction at energy $E_0$ provided that all the propagators
$G_0,P_\sigma,Q_\sigma$ and $T_\sigma$ are analytic at $E_0$.
\end{itemize}

The first point is easy. The RHS of (\ref{eq:decomp1d1},\ref{eq:decomp1d2})
has a singularity if $T_\uparrow(E,y_0) T_\downarrow(E,y_0) = 1$. i.e.
\begin{equation}
\delta_\uparrow(E,y_0)-\delta_\downarrow(E,y_0) = 0 \pmod{2\pi}.
\label{eq:qc1}
\end{equation}
This happens if and only if $E$ is an
eigenvalue of the Schr\" odinger operator $\op{H}$ since the two partial
solutions of the Schr\" odinger equation $\Psi^\sigma_{E y_0}(y),\,
\sigma = \uparrow,\downarrow$ (see
equation (\ref{eq:schr})) can be joined into the
continuous and differentiable (at SOS $y=y_0$) eigenfunction
\begin{equation}
\frac{\partial_y \Psi^\uparrow_{E y_0}(y_0)}{\Psi^\uparrow_{E y_0}(y_0)} =
\frac{\partial_y \Psi^\downarrow_{E y_0}(y_0)}{\Psi^\downarrow_{E y_0}(y_0)},
\label{eq:qc0}
\end{equation}
but $\Psi^\uparrow_{E y_0}(y_0) = (\cos\delta_\downarrow/
\cos\delta_\uparrow = \pm 1)\Psi^\downarrow_{E y_0}(y_0)$.
The quantization condition (\ref{eq:qc1},\ref{eq:qc0}) is independent of the
position of the SOS $y=y_0$, although in classically forbidden regions where
the wavenumber is imaginary
$k^2(E,y_0) < 0$ the phase-shift $\delta_\sigma(E,y_0)$ is also imaginary
number (plus an integer multiple of $\pi$)
and the SOS-SOS propagator $T_\sigma(E,y_0)$ becomes real with the magnitude
which is different from unity.

The second point is more elaborate. Let us define a
global solution of the Schr\" odinger equation
$\Ket{\Psi_{E y_0}}\in {\cal H}$
\begin{equation}
\Ket{\Psi_{E y_0}} =
\Ket{\Psi^\uparrow_{E y_0}}\pm\Ket{\Psi^\downarrow_{E y_0}}\\
\end{equation}
where we chose the sign ($\pm$ if $\half\delta$ is an even/odd multiple of
$\pi$) in order to make the eigenfunction $\Psi_{E_0 y_0}(y)$ continuous at
the SOS $y_0$. In the position representation
$\Psi_{E y_0}(\ve{x},y) = \Psi^\uparrow_{E y_0}(\ve{x},y) \pm
\Psi^\downarrow_{E y_0}(\ve{x},y)$, one term is always zero and the other is
not. If $E_0$ is an eigenenergy which satisfies quantization condition
(\ref{eq:qc1}) then the residuum of (\ref{eq:decomp1d2}) is
\begin{equation}
\Res\limits_{E=E_0}G(E,y^\prime,y) =
-\frac{\Psi_{E_0 y_0}(y^\prime)\Psi_{E_0 y_0}(y)}
{\partial_E \delta_\uparrow(E_0,y_0) - \partial_E \delta_\downarrow(E_0,y_0)}.
\label{eq:residuum}
\end{equation}
In order to calculate the energy derivative of the phase shift
$\partial_E \delta_\sigma(E,y)$ we should first study its dependence on $y$.
Note that any wavefunction $\Psi(y)$ which satisfies the Schr\" odinger
equation
\begin{equation}
(\partial^2_y + k^2(E,y))\Psi(y) = 0,\quad{\rm with}\quad
\Psi(y_\sigma) = 0
\label{eq:schr2}
\end{equation}
is proportional to $\Psi^\sigma_{E y_0}(y)$, for $\sigma y \ge \sigma y_0$
(see lemma \ref{lem:L1})
\begin{equation}
\Psi(y) = a\cos\left(k(E,y_0)(y-y_0) + \half\delta_\sigma(E,y_0)\right)
+ \Ord{|y-y_0|^2}.
\end{equation}
Let us differentiate this equation with respect to $y$, then set $y_0 = y$
and substitute $\Psi(y)$ back, giving
\begin{equation}
\partial_y \Psi(y) = -k(E,y)\tan\left(\half\delta(E,y)\right)\Psi(y)
\label{eq:derPsi}
\end{equation}
Now, differentiate the equation (\ref{eq:derPsi}) with respect to $y$,
use Schr\" odinger equation on the LHS and again (\ref{eq:derPsi}) on the RHS.
After cancelling the $\Psi(y)$ on both sides one obtains closed first order
nonlinear differential equation for the phase-shifts
\begin{equation}
\partial_y \delta = 2k - \frac{\partial_y k}{k}\sin(\delta),
\label{eq:derydelta}
\end{equation}
which is equivalent to the Riccati equation for the propagator $T_\sigma$
\begin{equation}
\partial_y T_\sigma = -2i\sigma k T_\sigma + \frac{\partial_y k}{2k}
(1 - T_\sigma^2).
\label{eq:deryT}
\end{equation}
The corresponding {\em initial} conditions are
\begin{equation}
\begin{array}{lll}
\delta_\sigma(E,y_\sigma) = \pi, & T_\sigma(E,y_\sigma) = -1 &
{\rm if}\;\; k^2(E,y_\sigma) \ge 0, \\
\delta_\sigma(E,y_\sigma) = n\pi-i\sigma\infty, & T_\sigma(E,y_\sigma) = 0 &
{\rm if}\;\; k^2(E,y_\sigma) < 0.
\end{array}
\label{eq:boundT}
\end{equation}
Then we define the so-called {\em incomplete normalization constants}
$N_\sigma(E,y)$
\begin{equation}
N_\sigma(E,y_0) = \int\limits_{y_0}^{y_\sigma} dy
\left[\Psi^\sigma_{E y_0}(y)\right]^2.
\label{eq:N1}
\end{equation}
Again we shall investigate the $y-$dependence of incomplete normalization
constants. Since $\Psi_{E y_0}(y)$ is again
proportional to some arbitrary wavefunction $\Psi(x)$ satisfying
Schr\" odinger {\em initial value problem} (\ref{eq:schr2})
(lemma \ref{lem:L1}) we can write
\begin{equation}
\Psi^\sigma_{E y_0}(y) = \Psi^\sigma_{E y_0}(y_0)\frac{\Psi(y)}{\Psi(y_0)}.
\label{eq:psiN}
\end{equation}
Use a substitution (\ref{eq:psiN}) in the definition of $N_\sigma(E,y_0)$
(\ref{eq:N1}), then move everything that does not depend on the
integration variable $y$ to the LHS and derive with respect to $y_0$, giving
$$
\partial_{y_0}\left[
\frac{\hbar^2 k(E,y_0)\Psi^2(y_0)}{4m \cos^2\left(\half
\delta_\sigma(E,y_0)\right)}N_\sigma(E,y_0)\right] =
-\Psi^2(y_0)
$$
After performing the differentiation and using the formulas
(\ref{eq:derPsi}) and (\ref{eq:derydelta}) for the derivative of $\Psi(y)$ and
$\delta_\sigma(E,y)$ with respect to $y$ one obtains a first order differential
equation for the incomplete normalization constants
\begin{eqnarray}
\partial_y N_\sigma + \frac{\partial_y k}{k}\cos(\delta_\sigma) N_\sigma &=&
-\frac{2m}{\hbar^2 k}(1 + \cos(\delta_\sigma)), \label{eq:deryN} \\
N_\sigma(E,y_\sigma) &=& 0.
\end{eqnarray}
Armed with formulas (\ref{eq:derydelta},\ref{eq:deryN}) we are ready to
determine the energy derivative of the phase-shift
$\partial_E \delta_\sigma(E,y)$ which is most valuable to us. It can be
uniquely determined from the differential equation which expresses the
uniqueness of the mixed second derivative
\begin{equation}
\partial_y(\partial_E \delta_\sigma) = \partial_E(\partial_y \delta_\sigma)
\label{eq:mix1}
\end{equation}
since the initial conditions are also known
\begin{equation}
\partial_E \delta_\sigma(E,y_\sigma) = 0.
\label{eq:mix1i}
\end{equation}
The author has used a heuristic ansatz
\begin{equation}
\partial_E \delta_\sigma = -N_\sigma
+ u(E,y)\cos(\delta_\sigma) + v(E,y)\sin(\delta_\sigma)
\end{equation}
and found that it is consistent with the system (\ref{eq:mix1},\ref{eq:mix1i})
iff $u = 0$ and $v = -\frac{m}{\hbar^2 k^2}$, so
\begin{equation}
\partial_E \delta_\sigma = -N_\sigma - \frac{m}{\hbar^2
k^2}\sin(\delta_\sigma).
\label{eq:derEdelta}
\end{equation}
Now, we can easily prove the rest of the decomposition formula with the
calculation of the residuum (\ref{eq:residuum}). If $E_0$ is an eigenenergy
then from the quantization condition (\ref{eq:qc1}) one has
$\sin(\delta_\uparrow(E_0,y_0)) = \sin(\delta_\downarrow(E_0,y_0))$ and
so the denominator of (\ref{eq:residuum})
\begin{eqnarray}
&&\partial_E \delta_\uparrow(E_0,y_0) - \partial_E \delta_\downarrow(E_0,y_0)
= -(N_\uparrow(E_0,y_0) - N_\downarrow(E_0,y_0)) = \nonumber \\
&&= - N(E_0,y_0) = -\int\limits_{y_\downarrow}^{y_\uparrow} dy
\left[\Psi_{E_0 y_0}(y)\right]^2
\label{eq:N2}
\end{eqnarray}
is just a negative {\em (complete) normalization constant} which makes the
residuum equal to the product of normalized eigenfunctions
\begin{equation}
\Res\limits_{E = E_0} G(E,y^\prime,y) = \Psi(y)\Psi(y^\prime),\quad
\Psi(y) = \frac{\Psi_{E_0 y_0}(y)}{\sqrt{N(E_0,y_0)}}.
\end{equation}
q.e.d.

\section*{IV. PROOF OF THE GENERAL CASE}
To prove the general decomposition formula (\ref{eq:decomp}) of Theorem
\ref{the:main} we shall use roughly the same procedure as for the 1-dim
case in previous section although few steps will be more lengthy now.

First we have to prove (see Point 1, Section III) that all poles of the
Green's operator (resolvent) $\op{G}(E)$ are due to singularities of
$(1 - \op{\T}_\downarrow(E,y_0)\op{\T}_\uparrow(E,y_0))$ in the energy
region where the propagator $\op{\T}_\sigma(E,y_0)$ is well defined.
The threshold energies for opening of new modes $E^\prime_n(y_0)$ should be
excluded so that the inverse $\op{K}^{-1/2}(E,y_0)$ which enters
into the definition of $\op{T}_\sigma(E,y_0)$ is well defined.

Any wavefunction $\Psi^\sigma_{E y_0}(\ve{x},y)$ which satisfies
Schr\" odinger equation above/below $y_0$,
$\Ket{\Psi^\sigma_{E y_0}}\in {\cal H}^\sigma_{E y_0}$ can be represented
with some $\ket{\psi}\in {\cal L}_{y_0}$ in a form
\begin{equation}
\Psi^\sigma_{E y_0}(\ve{x},y) = \Bra{\ve{x},y}\op{\Q}_\sigma(E,y_0)\ket{\psi}.
\end{equation}
If $E_0$ is such a pole, i.e. an eigenenergy of the Hamiltonian $\op{H}$,
then the corresponding eigenfunction $\Psi_1(\ve{x},y)$ and its normal
derivative $\partial_y \Psi_1(\ve{x},y)$ should be continuous on the
SOS ${\cal S}_{y_0}$. In general, the eigenenergy $E_0$ can sometimes be
degenerate, i.e. the corresponding eigenspace can be more than one, say,
$d$-dimensional spanned by the eigenfunctions $\Psi_n(\ve{x},y), n=1\ldots d$.
So, there should exist $2d$ SOS-states
$\ket{\sigma n} \in {\cal L}_{y_0},\,n=1\ldots d,\,\sigma=\uparrow,\downarrow$
such that
\begin{equation}
\Psi_n(\ve{x},y) = \left\{
\begin{array}{ll}
\Bra{\ve{x},y}\op{\Q}_\uparrow(E_0,y_0)\ket{\uparrow n}
&{\rm if}\;\; y \ge y_0,\\
\Bra{\ve{x},y}\op{\Q}_\downarrow(E_0,y_0)\ket{\downarrow n}
&{\rm if}\;\; y < y_0,
\end{array} \right. \label{eq:SOSes}
\end{equation}
with
\begin{equation}
\begin{array}{rcl}
\Bra{\ve{x},y_0}\op{\Q}_\uparrow(E_0,y_0)\ket{\uparrow n} &=&
\Bra{\ve{x},y_0}\op{\Q}_\downarrow(E_0,y_0)\ket{\downarrow n} \\
\partial_y \Bra{\ve{x},y}\op{\Q}_\uparrow(E_0,y_0)\ket{\uparrow n}
\vert_{y\searrow y_0} &=&
\partial_y \Bra{\ve{x},y}\op{\Q}_\downarrow(E_0,y_0)\ket{\downarrow n}
\vert_{y\nearrow y_0}.
\end{array}
\label{eq:cr}
\end{equation}
Continuity relations (\ref{eq:cr}) can be rewritten by using expansion
(\ref{eq:compexp}) to calculate the normal derivative
\begin{equation}
\partial_y \Bra{\ve{x},y}\op{\Q}_\sigma(E,y_0)\ket{\psi}
\vert_{\sigma y\searrow\sigma y_0} =
\sigma\frac{\sqrt{im}}{\hbar}\bra{\ve{x}}\op{K}^{1/2}(E,y_0)\left(1 -
\op{\T}_\sigma(E,y_0)\right)\ket{\psi}
\label{eq:derQ}
\end{equation}
and by using the completeness of position SOS-states $\ket{\ve{x}}$,
\begin{eqnarray}
\left(1 + \op{\T}_\uparrow(E_0,y_0)\right)\ket{\uparrow n} &=&
\left(1 + \op{\T}_\downarrow(E_0,y_0)\right)\ket{\downarrow n},
\label{eq:cr1}\\
\left(1 - \op{\T}_\uparrow(E_0,y_0)\right)\ket{\uparrow n} &=&
-\left(1 - \op{\T}_\downarrow(E_0,y_0)\right)\ket{\downarrow n}.
\label{eq:cr2}
\end{eqnarray}
Adding and subtracting equations (\ref{eq:cr1},\ref{eq:cr2}) one obtains
important relations
\begin{equation}
\op{\T}_\sigma(E_0,y_0) \ket{\sigma n} = \ket{-\sigma n},
\label{eq:applT1}
\end{equation}
which mean that the operator
$1 - \op{\T}_{-\sigma}(E_0,y_0)\op{\T}_\sigma(E_0,y_0)$ is singular with
$\ket{\sigma n}$ being the corresponding right null-vectors
\begin{equation}
\left(1 - \op{\T}_{-\sigma}(E_0,y_0)\op{\T}_\sigma(E_0,y_0)\right)
\ket{\sigma n} = 0. \label{eq:qc2}
\end{equation}
By deriving the SOS-quantization condition (\ref{eq:qc2}) we have proved the
point 1 completely, since the reasoning (\ref{eq:SOSes}-\ref{eq:qc2}) can
easily be reversed: existence of $d$-dim nulspace of operator
$1 - \op{\T}_{-\sigma}(E_0,y_0)\op{\T}_\sigma(E_0,y_0)$ implies the existence
of $d$-dim eigenspace of the Hamiltonian $\op{H}$.
Each $d$ SOS-states $\ket{\sigma n}, n=1\ldots d$ are linearly
independent, since $\sum_n c_n \ket{\sigma n} = 0$ would imply
$\sum_n c_n \Ket{\Psi_n} = 0$ due to linearity of relation (\ref{eq:SOSes}).
The null-space of $1 - \op{\T}_{-\sigma}(E_0,y_0)\op{\T}_\sigma(E_0,y_0)$
which is spanned by $\ket{\sigma n},\, n=1\ldots n$ is therefore also
$d$-dimensional.
\footnote{This is another proof of the Sturm-Liouville
theorem which forbids degeneracies in one dimension since then SOS-Hilbert
space ${\cal L}$ is 1-dimensional, and so $d$ cannot be greater than $1$.}
We shall soon need also the {\em complementary} SOS-representation of the
(complex conjugated) eigenfunctions $\Psi^*_n(\ve{x},y)$ which are in analogy
with (\ref{eq:SOSes}) provided by propagators $\op{\P}_\sigma(E_0,y_0)$.
There exists $2d$ SOS-states $\ket{\sigma n^*}\in {\cal L}_{y_0},\,
n = 1\ldots d,\,\sigma=\uparrow,\downarrow$, such that
\begin{equation}
\Psi_n^*(\ve{x},y) = \left\{
\begin{array}{ll}
\bra{\uparrow n^*}\op{\P}_\uparrow(E_0,y_0)\Ket{\ve{x},y}
&{\rm if}\;\; y \ge y_0,\\
\bra{\downarrow n^*}\op{\P}_\downarrow(E_0,y_0)\Ket{\ve{x},y}
&{\rm if}\;\; y < y_0.
\end{array} \right. \label{eq:SOSes2}
\end{equation}
Requiring continuity of eigenfunctions and their normal derivatives
results in a sequence of formulas completely analogous to
(\ref{eq:cr}-\ref{eq:cr2}) ending with
\begin{equation}
\bra{\sigma n^*}\op{\T}_\sigma(E_0,y_0) = \bra{-\sigma n^*},
\label{eq:applT2}
\end{equation}
and the complementary quantization condition
\begin{equation}
\bra{-\sigma n^*}
\left(1 - \op{\T}_{-\sigma}(E_0,y_0)\op{\T}_\sigma(E_0,y_0)\right) = 0.
\label{eq:qc3}
\end{equation}

We have still to prove (see Point 2, Section III)
that the residuum of the energy dependent Green's function
$\Res_{E = E_0}\Bra{\ve{q}^\prime}\op{G}(E)\Ket{\ve{q}}$ is equal to the
sum (in case of degeneracy) of products of orthonormalized eigenfunctions
$\sum_{n=1}^{d}\Psi_n(\ve{q}^\prime)\Psi_n^*(\ve{q})$
where complex conjugation should be used carefully since eigenfunctions
are no longer necessarily real. So let us assume that our sample
functions which span the eigenspace at $E_0$ are orthonormal
$\Braket{\Psi_n}{\Psi_l} = \delta_{nl}$. The following lemma will
be found very useful for calculating the residui of operator-valued
functions.
\begin{lemma}
\label{lem:L2}
Let $\op{A}$ be a singular operator $\op{A}\in Lin({\cal L})$ with
$d-$dimensional left and right null-space. If $\ket{L_n},\ket{R_n}\in {\cal L},
\,n = 1\ldots d$ span the left and right null space
${\cal N}_L = {\rm ker}(\op{A}^\dagger)$ and
${\cal N}_R = {\rm ker}(\op{A})$, respectively, then
\begin{equation}
\Res\limits_{z = 0} (z - \op{A})^{-1} = \sum\limits_{n,l=1}^d
{Inv}_{nl}[\braket{L_n}{R_l}] \ket{R_n}\bra{L_l}
\end{equation}
where ${Inv}_{nl}[c_{nl}]$ denotes the finite-dimensional-matrix inversion,
\begin{equation}
\sum\limits_{l=1}^d {Inv}_{nl}[\braket{L_n}{R_l}]\braket{L_l}{R_{n^\prime}}
=\sum\limits_{l=1}^d\braket{L_n}{R_l}{Inv}_{l n^\prime}
[\braket{L_l}{R_{n^\prime}}] = \delta_{n n^\prime} \label{eq:invmat}
\end{equation}
\end{lemma}
\noindent {\bf Proof} Let the left and right images be denoted by
${\cal I}_L = \op{A}^\dagger {\cal L}$ and
${\cal I}_R = \op{A} {\cal L}$, respectively.
We note that,
\begin{eqnarray}
{\cal I}_R &=& {\cal N}_L^\perp, \label{eq:IN1} \\
{\cal I}_L &=& {\cal N}_R^\perp. \label{eq:IN2}
\end{eqnarray}
Let us prove equation (\ref{eq:IN1}) first.\\
(i) Let $\ket{a}\in {\cal I}_R$, so $\exists\ket{b}\in {\cal L}$ such that.
$\ket{a} = \op{A}\ket{b}$. Then $\forall \ket{c}\in {\cal N}_L,\;
\braket{c}{a}=\bra{c}\op{A}\ket{b} = 0$, and therefore $\ket{a}\in
{\cal N}_L^\perp$. So, ${\cal I}_R \subseteq {\cal N}_L^\perp$.\\
(ii) Let $\ket{a}\in {\cal I}_R^\perp$, so $\forall\ket{b}\in{\cal L},\,
\bra{a}(\op{A}\ket{b}) = 0$, $\bra{a}\op{A} = 0$, and therefore $
\ket{a}\in {\cal N}_L$
So, ${\cal I}_R^\perp \subseteq {\cal N}_L = ({\cal N}_L^\perp)^\perp$.\\
Combining (i) and (ii) with the fact that ${\cal A} \subseteq {\cal B}
\iff {\cal A}^\perp \supseteq {\cal B}^\perp$ immediately gives (\ref{eq:IN1}).
Equation (\ref{eq:IN2}) need not be proved separately since it changes to
(\ref{eq:IN1}) after formal substitution of the operator under consideration
$\op{A}\rightarrow\op{A}^\dagger$.\\
Let $\ket{a}\in {\cal N}_L^\perp = {\cal I}_R$. So,
$\exists\lim_{z\rightarrow 0}(z-\op{A})^{-1}\ket{a} = \ket{c}$ since
$-\op{A}\ket{c} \in {\cal I}_R$. Then, of course
$\Res_{z=0}(z-\op{A})^{-1}\ket{a} = 0$.
Analogously for $\ket{b}\in {\cal N}_R^\perp = {\cal I}_L$,
we have $\Res_{z=0}\bra{b}(z-\op{A})^{-1} = 0$. The most general form
of the residuum of the resolvent which remains is
$\in Lin({\cal N}_L,{\cal N}_R)$,
\begin{equation}
\Res\limits_{z=0}(z - \op{A})^{-1} = \sum\limits_{n,l=1}^d
c_{nl} \ket{R_n}\bra{L_l}. \label{eq:res1}
\end{equation}
The unknown coefficients $c_{nl}$ can be determined by applying
(\ref{eq:res1}) on $\ket{R_{n^\prime}}$ and observing that $\op{A}$ is
null when restricted to ${\cal N}_R$, so
$(z- \op{A})^{-1}\ket{R_{n^\prime}} = z^{-1}\ket{R_{n^\prime}}$ and
$\Res\limits_{z=0}(z- \op{A})^{-1}\ket{R_{n^\prime}} = \ket{R_{n^\prime}}$.
Since vectors $\ket{R_{n^\prime}}$ are linearly independent, the
coefficients $c_{nl}$ satisfy
\begin{equation}
\sum\limits_{n,l=1}^d c_{nl}\braket{L_l}{R_{n^\prime}} = \delta_{n n^\prime},
\end{equation}
so they are the elements of the inverse martrix (\ref{eq:invmat}). q.e.d.

Let us return to physics. We have shown above (see equations (\ref{eq:qc2},
\ref{eq:qc3})) that the operator
$(1 - \op{\T}_{-\sigma}(E_0,y_0)\op{\T}_\sigma(E_0,y_0))$
has left and right null-space of dimension $d$ spanned by
$\ket{-\sigma n^*}$ and $\ket{\sigma n},\, n=1\ldots d$, respectively.
Let us now try to evaluate the residuum of its inverse
\begin{eqnarray}
\Res\limits_{E=E_0}\left(1 - \op{T}\right)^{-1}
&=& \Res\limits_{E=E_0}\left(1-\op{T}^0-(E-E_0)\partial_E\op{T}^0\right)^{-1} =
\nonumber\\
&=& -\Res\limits_{z=0}\left(z-(\partial_E\op{T}^0)^{-1}(1-\op{T}^0)\right)^{-1}
(\partial_E\op{\T}^0)^{-1} = \nonumber\\
&=& -\sum\limits_{n,l=1}^d
{Inv}_{nl}\left[\bra{-\sigma n^*}\partial_E\op{T}^0\ket{\sigma l}\right]
\ket{\sigma n}\bra{-\sigma l^*} = \label{eq:res2}\\
&=& -\sum\limits_{n,l=1}^d
{Inv}_{nl}\left[\bra{\uparrow n^*}\partial_E\op{T}_\uparrow^0
\ket{\uparrow l} + \bra{\downarrow n^*}\partial_E\op{T}_\downarrow^0
\ket{\downarrow l}\right]
\ket{\sigma n}\bra{-\sigma l^*} \nonumber
\end{eqnarray}
where shorthand notation
$\op{T} = \op{T}_{-\sigma}(E,y_0)\op{T}_\sigma(E,y_0),\,
\op{A}^0 = \op{A}(E_0,y_0)$ was introduced to avoid lengthy expressions.
We have used lemma $\ref{lem:L2}$ with
$\op{A} = (\partial_E\op{T}^0)^{-1}(1 - \op{T}^0),\,
\ket{R_n} = \ket{\sigma n},\,
\bra{L_n} = \bra{-\sigma n^*}\partial_E\op{T}^0$,
differentiated $\op{T}^0$ as a product, and in the end used transformations
(\ref{eq:applT1},\ref{eq:applT2}).

Therefore we have to study energy derivatives of the SOS-SOS propagators
$\partial_E\op{T}_\sigma(E,y)$. In analogy with the procedure in Section III,
we start by asking for $y$-dependence of $\op{T}_\sigma(E,y)$ first.
Take arbitrary one-side solution of the Schr\" odinger equation
$\Psi(\ve{x},y),\, \Ket{\Psi}\in{\cal H}^\sigma_{E y_1}$. By using
lemma \ref{lem:L1} we can calculate the values of the solution and its
normal derivative on any SOS ${\cal S}_y$ for $\sigma y_1 \le \sigma y
\le \sigma y_\sigma$ (i.e. everywhere $y_\downarrow \le y \le y_\uparrow$
since $y_1$ was arbitrary)
\begin{eqnarray}
\Psi(\ve{x},y) &=& \frac{\sqrt{-im}}{\hbar}\bra{\ve{x}}
\op{K}^{-1/2}(E,y)\left(1 + \op{T}_\sigma(E,y)\right)\ket{\psi}
\label{eq:valPsi},\\
\partial_y\Psi(\ve{x},y) &=& \sigma\frac{\sqrt{im}}{\hbar}\bra{\ve{x}}
\op{K}^{1/2}(E,y)\left(1 - \op{T}_\sigma(E,y)\right)\ket{\psi}.
\label{eq:derPsi2}
\end{eqnarray}
Noting the completeness of position states $\ket{\ve{x}}$ one can express
$\ket{\psi}$ from (\ref{eq:valPsi}) and insert it to (\ref{eq:derPsi2}) and
obtain very useful relation between the values and derivatives of the
wavefunction on the SOS in terms of the SOS-SOS propagator $\op{T}_\sigma$
\begin{equation}
\int\limits_{\cal S}d\ve{x}\partial_y\Psi(\ve{x},y)\ket{\ve{x}} =
i\sigma \op{K}^{1/2}(1 - \op{T}_\sigma)(1 + \op{T}_\sigma)^{-1}
\op{K}^{1/2}\int\limits_{\cal S}d\ve{x}\Psi(\ve{x},y)\ket{\ve{x}}.
\label{eq:dervalPsi}
\end{equation}
Differentiate the equation (\ref{eq:dervalPsi}) with respect to $y$ and
use the Schr\" odinger equation in a rather unusual form (which can be seen by
the definition \ref{def:K} of the operator $\op{K}(E,y)$)
\begin{equation}
\left(\partial_y^2 + \op{K}^2(\ve{x},y)\right)
\int\limits_{\cal S} d\ve{x}\Psi(\ve{x},y)\ket{\ve{x}} = 0
\end{equation}
on the LHS, and apply formula (\ref{eq:dervalPsi}) again on the RHS to
eliminate the normal derivatives $\partial_y\Psi(\ve{x},y)$.
We have used $\partial_y\ket{\ve{x}}_y\equiv 0$ which is
compatible with the definition of the $y-$derivative of an operator-valued
function which will be given in the next paragraph.
Since $\Psi(\ve{x},y)$ is arbitrary for a given $y$, the
vector $\int d\ve{x}\Psi(\ve{x},y)\ket{\ve{x}}$ runs over entire space
${\cal L}_y$, so the operators in front of it on the LHS and RHS should be
equal
$$
-\op{K}^2 = i\sigma\partial_y\left(\op{K}^{1/2}(1-\op{T}_\sigma)
(1+\op{T}_\sigma)^{-1}\op{K}^{1/2}\right) -
\op{K}^{1/2}(1-\op{T}_\sigma)(1+\op{T}_\sigma)^{-1}\op{K}
(1-\op{T}_\sigma)(1+\op{T}_\sigma)^{-1}\op{K}^{1/2}.
$$
After formal algebraic manipulation and with the definition of few new
symbols \begin{eqnarray}
[\op{A},\op{B}]_\pm &=& \op{A}\op{B} \pm \op{B}\op{A}, \\
\op{K}_\pm &=& \half \left[\partial_y (\op{K}^{1/2}),\op{K}^{-1/2}\right]_\pm,
\end{eqnarray}
where $\op{K}^{-1/2} = (\op{K}^{1/2})^{-1}$,
one obtains closed first order operator Riccati equation for the
SOS-SOS propagator
\begin{equation}
\partial_y\op{T}_\sigma =
-i\sigma [\op{K},\op{T}_\sigma]_+ + [\op{K}_-,\op{T}_\sigma]_- +
\op{K}_+ - \op{T}_\sigma\op{K}_+\op{T}_\sigma
\label{eq:deryT2}
\end{equation}
with the initial condition
\begin{equation}
\op{T}_\sigma(E,y_\sigma) = 0
\label{eq:iniT}
\end{equation}
assuming that on the boundary $y=y_\sigma$ all SOS-wavenumbers
$k_n(E,y_\sigma)$ are imaginary.
(If not, they can be made such if the (fictitious) hard walls on the boundaries
$y=y_\sigma$, which ensure the Dirichlet boundary conditions,
are infinitesimally smoothed.)

A remark is in order concerning the derivation in the
last paragraph:
The derivative of the propagator $\op{T}_\sigma(E,y)$ with respect to $y$ is
perfectly well defined, although $\op{T}_\sigma(E,y)$ and
$\op{T}_\sigma(E,y+dy)$ act in different Hilbert spaces, since the latter are
isomorphic. So for any such operator-valued function of $y$
$\op{A}(y)\in Lin({\cal L}_y)$, one should define
\begin{equation}
\partial_y\op{A}(y) = \lim\limits_{\epsilon\rightarrow 0}
\frac{1}{\epsilon}\left(
\op{I}_{y,y+\epsilon}\op{A}(y+\epsilon)\op{I}_{y+\epsilon,y} -
\op{A}(y)\right).
\end{equation}

We shall also need the $y$-derivatives of the SOS-CS-SOS propagators
$\partial_y\op{\Q}_\sigma(E,y)$ and $\partial_y\op{\P}_\sigma(E,y)$.
Let us start by calculating the values of
$\partial_y\op{\Q}_\sigma(E,y)\ket{\psi}$ on the SOS ${\cal S}_y$ for
any $\ket{\psi}\in{\cal L}_y$.
\begin{eqnarray}
\Bra{\ve{x},y^\prime}\partial_y\op{\Q}_\sigma(E,y)\ket{\psi}
\vert_{\sigma y^\prime\searrow\sigma y}
&=& \partial_y\left(\Bra{\ve{x},y}\op{\Q}_\sigma(E,y)\ket{\psi}\right)
- \partial_{y^\prime}\Bra{\ve{x},y^\prime}\op{\Q}_\sigma(E,y)\ket{\psi}
\vert_{\sigma y^\prime\searrow\sigma y} = \nonumber\\
&=& \frac{\sqrt{im}}{\hbar}\bra{\ve{x}}\left[
\partial_y\left(\op{K}^{-1/2}(1 + \op{T}_\sigma)\right)
- i\sigma K^{1/2}(1 - \op{T}_\sigma)\right]\ket{\psi} = \nonumber\\
&=& \Bra{\ve{x},y}\op{\Q}_\sigma\left[-i\sigma\op{K} - \op{K}_-
- \op{K}_+\op{T}_\sigma\right]\ket{\psi}, \label{eq:boundderQ}
\end{eqnarray}
where the equations (\ref{eq:valQ},\ref{eq:derQ}) and (\ref{eq:deryT2})
and some operator algebra were applied.
Note that the derivative of the solution of the Schr\" odinger equation
$\Psi(\ve{x},y^\prime) = \Bra{\ve{x},y^\prime}\op{\Q}_\sigma(E,y)
\ket{\psi}$ with
respect to smooth external parameter $y$ is again solution of the
Schr\" odinger equation, $\partial_y\op{\Q}_\sigma(E,y)\ket{\psi}\in
{\cal H}^\sigma_{E y}$. The solution of the {\em linear} Schr\" odinger
equation is a unique and {\em linear} function of boundary conditions, so the
linear relation for boundary condition (\ref{eq:boundderQ}) holds
also globally (and since ket $\ket{\psi}$ is arbitrary it can be omitted)
\begin{equation}
\partial_y \op{\Q}_\sigma = \op{\Q}_\sigma\left[
-i\sigma\op{K} - \op{K}_- - \op{K}_+\op{\T}_\sigma\right].
\label{eq:deryQ}
\end{equation}
One can completely analogously derive similar equation for another propagator
$\op{\P}_\sigma(E,y)$
\begin{equation}
\partial_y \op{\P}_\sigma = \left[
-i\sigma\op{K} + \op{K}_- - \op{\T}_\sigma\op{K}_+\right]\op{\P}_\sigma.
\label{eq:deryP}
\end{equation}
In analogy with 1-dimensional case (although this
definition slightly differs from that in Section III) we define also
an {\em incomplete normalization operator}
$\op{N}_\sigma(E,y)\in Lin({\cal L}_y),$
\begin{eqnarray}
\op{N}_\sigma(E,y) &=& \sigma\int\limits_{\cal
S}d\ve{x}\int\limits_y^{y_\sigma}
dy^\prime \op{\P}_\sigma(E,y)\Ket{\ve{x},y^\prime}\Bra{\ve{x},y^\prime}
\op{\Q}_\sigma(E,y) = \label{eq:intN} \\
&=& \op{\P}_\sigma(E,y)\op{Q}_\sigma(E,y). \label{eq:ordN}
\end{eqnarray}
We are again interested in the derivative of the incomplete normalization
operator with respect to $y$.
So far we have ignored the discontinuities of $\Bra{\ve{x},y^\prime}
\op{\Q}_\sigma(E,y)$ and $\op{\P}_\sigma(E,y)\Ket{\ve{x},y^\prime}$
at $y^\prime = y$ by always approaching $y^\prime\rightarrow y$ from the
proper side, so we have to take representation $(\ref{eq:intN})$ to account for
the discontinuities, and calculate
\begin{eqnarray*}
\partial_y\op{N}_\sigma(E,y) &=&
\left(\partial_y\op{\P}_\sigma(E,y)\right)\op{\Q}_\sigma(E,y) +
\op{\P}_\sigma(E,y)\partial_y\op{\Q}_\sigma(E,y) -\\
&-& \sigma\int\limits_{\cal S}
d\ve{x}\op{\P}_\sigma(E,y)\Ket{\ve{x},y}\Bra{\ve{x},y}\op{\Q}_\sigma(E,y).
\end{eqnarray*}
The remaining integral over the SOS can be calculated by means of initial data
(\ref{eq:valQ},\ref{eq:valP})
$$
\int\limits_{\cal S}
d\ve{x}\op{\P}_\sigma(E,y)\Ket{\ve{x},y}\Bra{\ve{x},y}\op{\Q}_\sigma(E,y)
= -\frac{im}{\hbar^2}(1 + \op{\T}_\sigma)\op{K}^{-1}(1 + \op{\T}_\sigma),
$$
so after applying formulas (\ref{eq:deryQ},\ref{eq:deryP}) one can write
the differential system for the incomplete normalization operator
\begin{eqnarray}
\partial_y\op{N}_\sigma &=& -i\sigma[\op{K},\op{N}_\sigma]_+ +
[\op{K}_-,\op{N}_\sigma]_- - \op{\T}_\sigma\op{K}_+\op{N}_\sigma -
\op{N}_\sigma\op{K}_+\op{\T}_\sigma + \nonumber \\
&+& i\sigma\frac{m}{\hbar^2}(1 + \op{T}_\sigma)\op{K}^{-1}(1 + \op{T}_\sigma),
\label{eq:deryN2}\\
\op{N}_\sigma(E,y_\sigma) &=& 0,
\end{eqnarray}
which is linear in $\op{N}_\sigma$, but depends again quadratically on
$\op{T}_\sigma$ (compare with (\ref{eq:deryT2})).
Now we have the tools to ask for the derivative of the SOS-SOS propagator
with respect to the energy $\partial_E\op{\T}_\sigma$. It can be determined
as a unique solution of the first-order differential system
\begin{eqnarray}
\partial_y\left(\partial_E\op{\T}_\sigma\right) &=&
\partial_E\left(\partial_y\op{\T}_\sigma\right), \label{eq:mix2} \\
\partial_E\op{\T}_\sigma(E,y_\sigma) &=& 0, \label{eq:imix2}
\end{eqnarray}
which is the obvious requirement for the uniqueness of the mixed second
derivative. The initial condition (\ref{eq:imix2}) is just an energy
derivative of the initial condition (\ref{eq:iniT}).
The author has guessed the solution in analogy with 1-dimensional case
(Section III)
\begin{equation}
\partial_E\op{\T}_\sigma = -\op{N}_\sigma + \frac{m}{2\hbar^2}\left(
\op{K}^{-2} - \op{\T}_\sigma\op{K}^{-2}\op{\T}_\sigma\right).
\label{eq:derET}
\end{equation}
The reader can verify the consistency of (\ref{eq:derET}) by inserting it into
(\ref{eq:mix2}) and applying the formulas (\ref{eq:deryT2},\ref{eq:deryN2}) and
the formulas
\begin{eqnarray}
\partial_E\op{K}^p &=& \frac{m}{\hbar^2}p\op{K}^{p-2}, \\
\partial_E\op{K}_- &=& \frac{m}{2\hbar^2}[\op{K}^{-2},\op{K}_+]_- , \\
\partial_E\op{K}_+ &=& \frac{m}{2\hbar^2}\left([\op{K}^{-2},\op{K}_-]_- +
\partial_y(\op{K}^{-2})\right),
\end{eqnarray}
which can be derived directly from the definitions and
with a bit of operator algebra, of course.

Now we can finish the calculation of the residuum (\ref{eq:res2}).
First we use our new result (\ref{eq:derET}) and the formulae
(\ref{eq:applT1},\ref{eq:applT2}) and (\ref{eq:SOSes},\ref{eq:SOSes2})
with definition (\ref{eq:intN}) to calculate the matrix elements
\begin{eqnarray*}
\bra{\uparrow n^*}\partial_E\op{T}_\uparrow^0
\ket{\uparrow l} + \bra{\downarrow n^*}\partial_E\op{T}_\downarrow^0
\ket{\downarrow l} &=&
-\bra{\uparrow n^*}\op{N}_\uparrow^0\ket{\uparrow l}
-\bra{\downarrow n^*}\op{N}_\downarrow^0\ket{\downarrow l} = \\
&=&-\int\limits_{\cal S}d\ve{x}\left(
\int\limits_y^{y_\uparrow} dy^\prime +
\int\limits_{y_\downarrow}^y dy^\prime\right)
\Psi_n^*(\ve{x},y^\prime)\Psi_l(\ve{x},y^\prime) = \\
&=&-\delta_{nl},
\end{eqnarray*}
since the eigenstates $\Ket{\Psi_n}$ are supposed to be orthonormal
$\Braket{\Psi_n}{\Psi_l} = \delta_{nl}$.
The residuum (\ref{eq:res2}) is therefore very simple
\begin{equation}
\Res\limits_{E=E_0}(1-\op{T})^{-1} =
\sum\limits_{n=1}^d \ket{\sigma n}\bra{-\sigma n^*}.
\end{equation}
The calculation of the residuum of the general Green's function
(\ref{eq:decomp}) is now straightforward
\begin{eqnarray}
\Res\limits_{E=E_0}\Bra{\ve{q}^\prime}\op{G}(E)\Ket{\ve{q}} &=&
\sum\limits_{\sigma,n}
\Bra{\ve{q}^\prime}\op{\Q}_\sigma^0 \ket{\sigma n}\bra{-\sigma n^*}
\op{\P}_{-\sigma}^0\Ket{\ve{q}} + \nonumber \\
&+& \sum\limits_{\sigma,n}
\Bra{\ve{q}^\prime}\op{\Q}_\sigma^0 \ket{\sigma n}\bra{-\sigma n^*}
\op{T}_{-\sigma}^0\op{\P}_{\sigma}^0\Ket{\ve{q}} = \nonumber \\
&=& \sum\limits_{\sigma,\sigma^\prime,n}
\Bra{\ve{q}^\prime}\op{\Q}_{\sigma^\prime}^0 \ket{\sigma^\prime n}
\bra{\sigma n^*}\op{\P}_\sigma^0\Ket{\ve{q}} = \nonumber \\
&=& \sum\limits_n \Psi_n(\ve{q}^\prime)\Psi_n^*(\ve{q}).
\end{eqnarray}
This completes the proof of the general decomposition formula.

\section*{V. DISCUSSION AND CONCLUSIONS}

In this paper I have presented and proved the decomposition of the resolvent
$\op{G}(E) = (E - \op{H})^{-1}$ (formula (\ref{eq:decomp})) in terms of four
newly defined energy-dependent propagators from and/or to Hilbert space of
complex-valued functions over $f$-dim configuration space (CS) (kets and
bras with angle brackets) to and/or from
Hilbert space of complex-valued functions over $(f-1)$-dim configurational
surface of section (SOS) (kets and bras with curly brackets)
which have the following clear physical interpretations
\begin{itemize}
\item $\Bra{\ve{q}^\prime}\op{G}_0(E,y)\Ket{\ve{q}}$ is a quantum
probability amplitude
that the system propagates from point $\ve{q}$ to point $\ve{q}^\prime$ at
energy $E$ without hitting the SOS labeled by $y$.
\item $\bra{\ve{x}}\op{\P}_\sigma(E,y)\Ket{\ve{q}}$ is a probability amplitude
that the system propagates from point $\ve{q}$ above (if $\sigma=\uparrow$) or
below (if $\sigma=\downarrow$) SOS to the point $\ve{x}$ on the SOS at energy
$E$ and without hitting the SOS in between.
\item $\Bra{\ve{q}}\op{\Q}_\sigma(E,y)\ket{\ve{x}}$ is a probability amplitude
that the system propagates from point $\ve{x}$ on the SOS to the point $\ve{q}$
in the CS above (if $\sigma=\uparrow$) or below (if $\sigma=\downarrow$) SOS
at energy $E$ without hitting the SOS in between.
\item $\bra{\ve{x}^\prime}\op{\T}_\sigma(E,y)\ket{\ve{x}}$ is a probability
amplitude that the system propagates from point $\ve{x}$ on the SOS to the
point $\ve{x}^\prime$ on the SOS through the upper (if $\sigma=\uparrow$) or
lower (if $\sigma=\downarrow$) part of CS with respect to the SOS and at
energy $E$ without hitting the SOS in between.
\end{itemize}
All these propagators are
analytic (holomorphic) in the upper half energy plane, $\im E > 0$.
By expressing them in terms of the resolvent of the related scattering
Hamiltonian $\cite{P94a}$ one can show that they can have typically a
{\em finite number of poles} on the {\em real energy-axis} or they have
even no poles on the entire real energy-axis if a given SOS ${\cal S}_y$
is being crossed by every classical trajectory, heuristically speaking
(or the corresponding scattering problem, whoose S-matrix is just our
propagator $\op{\T}_\sigma(E,y)$ \cite{P94a}, has no bound states,
rigorously speaking).

Using this interpretation one can arrive at the decomposition formula
(in expanded form $(\ref{eq:decomp2})$) heuristically, just by using
basic postulates of quantum mechanics about addition and
multiplication of probability amplitudes and knowing that
typical quantum path is continuous (but not differentiable) \cite{FH65}
(this is why the signs $\uparrow$ and
$\downarrow$ are alternating in the decomposition formula (\ref{eq:decomp2}).)

In case where there is a time-reversal symmetry, the quantum Poincar\' e
(SOS-SOS) map is symmetric in position representation \cite{P94b,RS94}
\begin{equation}
\bra{\ve{x}^\prime}\op{\T}_\sigma\ket{\ve{x}} =
\bra{\ve{x}}\op{T}_\sigma\ket{\ve{x}^\prime}.
\end{equation}

It can also be shown \cite{P94b,RS94} that if the SOS-SOS operator is
represented by a finite-dimensional matrix
$T^\sigma_{n^\prime n} = \bra{n^\prime}\op{\T}_\sigma\ket{n}$
in a truncated basis $\{\ket{n};k_n^2 > 0\}$ where we discard all
evanescent (closed) SOS-eigenmodes with imaginary wavenumbers then the matrix
$T^\sigma_{n^\prime n}$ is unitary.

The quantum surface of section is expected to be of great practical (numerical)
value since the dimension of the truncated matrices of the SOS-SOS propagators
(which is $\Ord{\hbar^{1-f}}$) can be orders of magnitude smaller than
the dimension of the truncated matrix of the full Hamiltonian
(which is $\Ord{\hbar^{-f}}$). Thus the practical algorithms based on quantum
SOS method for calculation of energy spectra can be developed which are orders
of magnitude faster than the existing ones.

The determination of the exact
SOS-SOS propagator (unlike the semiclassical one as in \cite{B92})
is still quite a difficult task. Solving the Riccati equations
(\ref{eq:deryT},\ref{eq:deryT2}) of the propagator flow through the
family of surfaces of section is quite elaborate, even numerically.
One would wish to have a direct method which has been so far successful only
for
special systems, like free particle on a ``trombone'' attached to a
parallelogram\cite{G93}, or Sinai billiard\cite{SS93}. The author succeeded
to find explicit procedure for calculation of the SOS-SOS propagator for
the so-called {\em semi-separable systems},
which are separable for $y > y_0$, and
for $y < y_0$, but they are possibly discontinuous at $y=y_0$.
The method has already been applied to one such 2-dim system with generic
nonintegrable classical dynamics \cite{P94c} and energy levels with sequential
quantum number around 20 million were easily calculated
(1 - 3 minutes of supercomputer (Convex C3860) CPU time per level).

Some of the open problems of the subject are: (i) developing the
phase-space version of the present theory e.g. in the Wigner-Weyl language,
(ii) developing practical methods for evaluating exact
SOS-propagators in concrete examples, (iii) establishing the connection between
the present formalism and path integrals with geometrical constraints which
are still to be defined for this purpose, whereas (iv) generalization of the
present theory, e.g. to cases of general magnetic fields where the component
of the momentum perpendicular to SOS does no longer enter the Hamiltonian only
quadratically but also linearly, and (v) study of the semiclassical limit
were already acomplished in \cite{P94b}.

In the present theory the algebra of the generators of parallel motions
(with respect to the SOS) need not be just the operator algebra
spanned by $(\ve{x},\ve{p}_x)$. Indeed an obvious generalization is possible
by considering an arbitrary Lie algebra instead, since we always refer just
to inside Hamiltonian  $\op{H}^\prime$ (see equation (\ref{eq:qH})) as a whole
and never to $\op{\ve{x}}$ and $\op{\ve{p}}_x$ separately.
For example, instead of ordinary position and momentum, $\ve{x}$ and
$\ve{p}_x$, we could have angular momentum or spin coordinates $J_1,J_2,J_3$.

\section*{ACKNOWLEDGMENTS}

I am grateful to Professor Marko Robnik for proposing the semiclassical
version of this project which lead me to the present work, for the
discussions on the subject, and for reading the manuscript.
The financial support by the Ministry of Science
and Technology of the Republic of Slovenia is gratefully acknowledged.

\vfill

\vfill
\newpage

\section*{Figure captions}
\bigskip
\bigskip

\noindent {\bf Figure 1} Illustration of one of the terms in the expanded form
of the decomposition formula for the energy dependent propagator
(\ref{eq:decomp2}), namely $\op{\Q}_\downarrow
\op{\T}_\uparrow\op{\T}_\downarrow\op{\P}_\uparrow$. A typical
continuous (but nondifferentiable) path which contributes to this term is shown
and the corresponding probability amplitudes are marked. The latter have
to be multiplied and sumed (integrated) over the intersection points
$\ve{x},\ve{x}^\prime$ and $\ve{x}^{\prime\prime}$ in order to yield the
probability amplitude for the entire term, namely the probability
amplitude to propagate from point $\ve{q}$ to point $\ve{q}^\prime$ at
definite energy $E$ and crossing the SOS exactly three times
$\Bra{\ve{q}^\prime}\op{\Q}_\downarrow\op{\T}_\uparrow\op{\T}_\downarrow
\op{\P}_\uparrow\Ket{\ve{q}}$.
\bigskip
\bigskip

  \setlength{\unitlength}{0.0080mm}
  \newcommand{\bpoint}[1]{\put#1{\circle*{150}}}
  \newcommand{\opoint}[1]{\put#1{\circle{150}}}
  \noindent
  \begin{picture}(12000,8100)
   \thinlines
   \put(0,0){\line(1,0){12000}}
   \put(12000,0){\line(0,1){8100}}
   \put(12000,8100){\line(-1,0){12000}}
   \put(0,8100){\line(0,-1){8100}}
   \put(1000,2700){\line(3,1){2108}}
   \put(3000,3367){\line(4,1){3092}}
   \put(6000,4117){\line(5,1){3059}}
   \put(9000,4717){\line(6,1){2028}}
   \put(10000,5300){SOS}
   \thicklines
   \put(1000,7500){\line(3,-1){1650}}
   \put(2650,6950){\line(-1,-4){587.5}}
   \put(2062.5,4600){\line(-1,0){1162.5}}
   \put(900,4600){\line(1,-1){2000}}
   \put(2900,2600){\line(0,-1){1400}}
   \put(2900,1200){\line(4,-1){2000}}
   \put(4900,700){\line(2,1){1200}}
   \put(6100,1300){\line(1,5){500}}
   \put(6600,3800){\line(-1,6){500}}
   \put(6100,6800){\line(2,-1){1000}}
   \put(7100,6300){\line(1,1){1500}}
   \put(8600,7800){\line(1,-6){900}}
   \put(9500,2400){\line(2,-1){1000}}
   \put(10500,1900){\line(-2,-1){1000}}
   \bpoint{(1000,7500)}
   \put(950,7050){$\ve{q}$}
   \bpoint{(9500,1400)}
   \put(9450,1000){$\ve{q}^\prime$}
   \opoint{(2360,3156)}
   \put(2050,2700){$\ve{x}$}
   \opoint{(6527,4223)}
   \put(6050,3730){$\ve{x}^\prime$}
   \opoint{(9117,4735)}
   \put(8550,4230){$\ve{x}^{\prime\prime}$}
   \put(80,5000){$\bra{\ve{x}}\op{\P}_\uparrow\Ket{\ve{q}}$}
   \put(3450,1330){$\bra{\ve{x}^\prime}\op{\T}_\downarrow\ket{\ve{x}}$}
   \put(6450,5700){$\bra{\ve{x}^{\prime\prime}}\op{\T}_\uparrow
    \ket{\ve{x}^\prime}$}
   \put(9560,2600){$\Bra{\ve{q}^\prime}\op{Q}_\downarrow
    \ket{\ve{x}^{\prime\prime}}$}
  \end{picture}

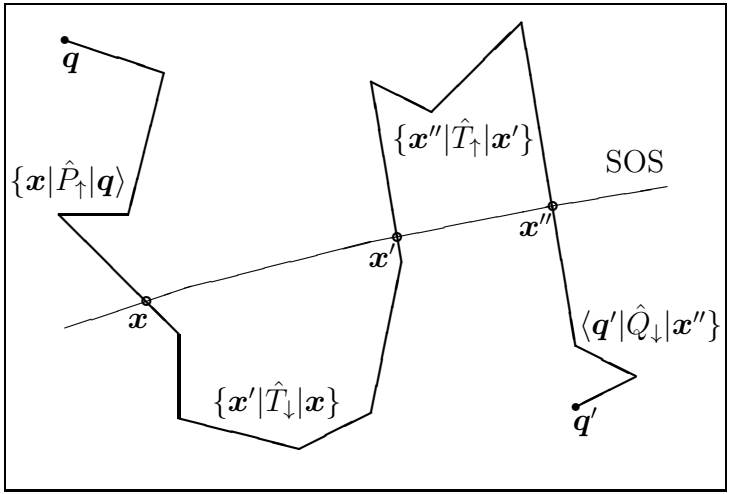
\end{document}